\documentclass[aps,prl,twocolumn,showpacs,superscriptaddress]{revtex4}

\usepackage{epsfig}
\usepackage{amsmath}

\newcommand{\todo}[1]{}

\begin{document}

\title{Direct control of the tunnel splitting in a one--electron
double quantum dot}

\author{A.~K.~H\"uttel}
\author{S. Ludwig}
\affiliation{Center for NanoScience and Department f\"ur Physik,
Ludwig--Maximilians--Universit\"at, Geschwister--Scholl--Platz 1,
80539~M\"unchen, Germany,}

\author{K.~Eberl}
\affiliation{Max-Planck-Institut f\"ur Festk\"orperforschung,
Heisenbergstra{\ss}e 1, 70569 Stuttgart, Germany.}

\author{J.~P.~Kotthaus}
\affiliation{Center for NanoScience and Department f\"ur Physik,
Ludwig--Maximilians--Universit\"at, Geschwister--Scholl--Platz 1,
80539~M\"unchen, Germany,}

\date{January 3, 2005}
 
\begin{abstract}
Quasi--static transport measurements are employed on a laterally defined
tunnel--coupled double quantum dot. A nearby quantum point contact allows us to
track the charge as added to the device.
If charged with only up to one electron, the low--energy spectrum of the double
quantum dot is characterized by its quantum mechanical interdot tunnel
splitting. We directly measure its magnitude by utilizing
particular anticrossing features in the
stability diagram at finite source--drain bias.
By modification of gate voltages defining the confinement potential as well
as by variation of a perpendicular magnetic field we
demonstrate the tunability of the coherent tunnel coupling.
\end{abstract}

\pacs{
73.21.La,    
73.23.Hk,    
73.20.Jc     
}

\maketitle

Recent works have shown spectacular advancements regarding the control
over single electrons trapped in semiconductor based quantum
dots (QD)~\cite{dots, ciorga, cmm-spinblock, spinfilter}. Electronic
states in separate QDs can be coupled, resulting in
delocalized and spatially coherent ``molecular
modes''~\cite{robert,alex,hayashi,dqdtarucha}. QDs thus lend themselves as prospective
building blocks for qubits, the elementary units of the proposed
quantum computer. Models for QD--based qubits include e.g. 
the use of a single electronic spin in one
QD~\cite{lossdivi,review}. Alternatively, the position of a single
electronic charge within a coupled double quantum dot
(DQD) has been proposed~\cite{ieee_blick,jjap_vdw}. For
both approaches, precise control of the coupling between nearby QDs
is of paramount importance.
We report on quasi--static measurements allowing direct 
determination of the interdot tunnel splitting
of a strongly coupled DQD charged with up to one electron. Control over the
tunnel coupling via gate voltages or magnetic field is demonstrated. 

\begin{figure}[tb]\begin{center}
\epsfig{file=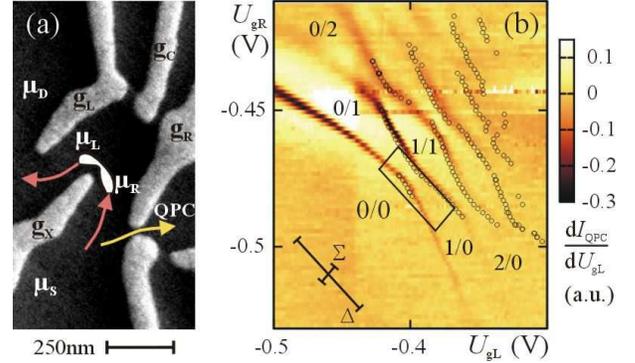, width=8cm}
\end{center}
\vspace*{-0.4cm}
\caption{
(Color online)
(a) SEM micrograph of the gate electrodes used for defining a DQD
and a QPC. Approximate geometry of the DQD (white area) and electron flow
(arrows) are indicated.
(b) Stability diagram of the DQD plotting the transconductance  ${{\rm d}
I_\mathrm{QPC}}/{{\rm d} U_\mathrm{gL}}$ through the QPC.
For clarity a background signal has been substracted.
The black circles indicate positions of high current through the DQD.
The bars ($\Sigma, \Delta$), each corresponding to $1\,$meV, illustrate the
axis directions, the box the plot range of Figs.~\ref{fig2} and \ref{fig3}.
}
\label{fig1}
\end{figure}

Our measurements are performed using an epitaxial AlGaAs/GaAs
heterostructure that forms a two--dimensional electron system (2DES)
$120\,\mathrm{nm}$ below the crystal surface with a carrier sheet
density of $n_{\rm s} = 1.8 \times 10^{15}\,\mathrm{m}^{-2}$ and an
electron mobility of $\mu=75\,\mathrm{m}^{2}/\mathrm{Vs}$, both
measured at $4.2\,$K.  The 2DES temperature is estimated to be $T_{\rm
2DES}\simeq100\,\mathrm{mK}$.  We use a lock--in frequency of $840
\,\mathrm{Hz}$ (Fig.~\ref{fig1}) or $680 \,\mathrm{Hz}$
(Figs.~\ref{fig3} and \ref{fig4}).

Fig.~\ref{fig1}(a)
displays an electromicrograph of the gates on the surface of our
heterostructure. Application of negative dc--voltages to these electrodes allows
definition of a DQD as well as a quantum point contact (QPC) in the 2DES.
Our layout is based on a gate geometry 
introduced by Ciorga
{\it et al.}~\cite{ciorga}. It allows a measurable single
electron tunneling (SET) current through the QD even in the limit
of only one trapped electron. To elongate an originally single QD and 
finally split it into a double well potential we increase the absolute values of
the dc--voltages $|U_{\rm gC}|$ and $|U_{\rm gX}|$ of the center gates marked
accordingly in Fig.~\ref{fig1}(a). Simultaneously we decrease $|U_{\rm gL}|$ and
$|U_{\rm gR}|$ on the side gates in order to keep the overall charge of the
system constant. The approximately resulting serial DQD is sketched in Fig.~\ref{fig1}(a) in
bright tone. 

The stability diagram in Fig.~\ref{fig1}(b) displays the differential
transconductance $G_{\rm QPC}={\rm d}I_{\rm QPC}/{\rm d}U_{\rm gL}$
through the nearby QPC in dependence on the dc--voltages
applied to gates $\rm g_L$ and $\rm g_R$. The dark lines in the gray scale plot depict
reduced transconductance corresponding to discrete charging events while otherwise the
electron number in the DQD is constant~\cite{field}.
These lines clearly form a honeycomb structure as expected for a
DQD~\cite{elzerman,dodoreview,hofmann}.
The lack of charging events in the area
marked by $0/0$ in Fig.~\ref{fig1}(b) 
implies that here our DQD is free of extra
charges. Note that this area is bordered on its upper and right side by a
distinct line of low transconductance proving that our resolution is high
enough to detect all charging events for the shown gate voltages.
We mark the area of the plot for which only the right QD is charged by one 
electron with $0/1$, for which one electron is trapped in each QD with $1/1$,
and so on. The smooth charge redistribution from configuration $0/1$ to $1/0$ contrasts
results for weak interdot coupling~\cite{elzerman, petta}. A second hint at
strong coupling is
the lack of sharp corners of the lines of reduced transconductance
at triple points
in the stability diagram, where three charge configurations are possible.

The open circles in Fig.~\ref{fig1}(b) depict local maxima of  
current $I_{\rm DQD}$ through the serial DQD recorded in linear response
($U_{\rm SD}=50\,\mu\mathrm{V}$). For weak interdot coupling we expect
finite current through this system only at triple points
in the stability diagram.
In contrast, we also find current along
parts of the configuration boundaries away from triple points.
This suggests 
that the electron can lower its orbital energy by forming a delocalized state, 
as well indicating strong interdot tunnel coupling.  
The configuration boundaries obtained show excellent
agreement with the QPC measurement. Note that we can
perform direct current measurements through the DQD even if it is charged with only
up to one electron.

Starting from the limit of a weakly coupled DQD we will now outline a model
including coherent interdot tunnel coupling, describing
the static transport properties at finite source--drain bias.
Comparison with measurements
demonstrates our ability to detect and tune the tunnel
splitting of the single electron states in a DQD.

\begin{figure}[tb]\begin{center}
\epsfig{file=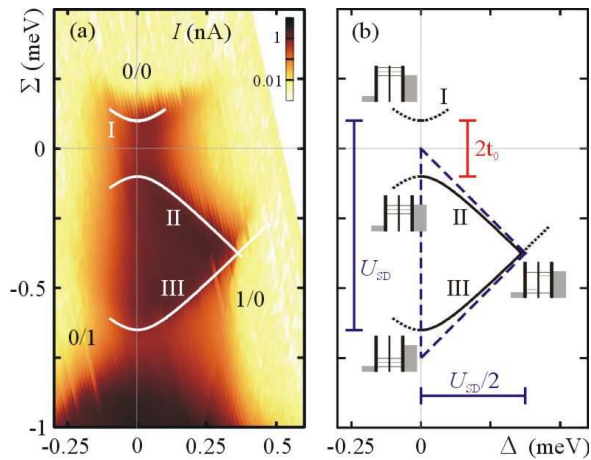, height=6cm}
\end{center}
\vspace*{-0.4cm}
\caption{
(Color online)
(a) 
Measurement of SET current through a one--electron DQD at $U_{\rm
SD}=-0.75\,\mathrm{mV}$ and $U_{\rm gC}= -1.47\,\mathrm{V}$
(logarithmic scale), including
model lines for strong interdot coupling (see (b)).
(b) 
Corresponding model expectations for SET transport features at finite
source--drain bias in the two cases of weak (dashed lines) versus strong (solid and
dotted lines) interdot coupling.
}
\label{fig2}
\end{figure}

\begin{figure}[tb]
\begin{center}
\epsfig{file=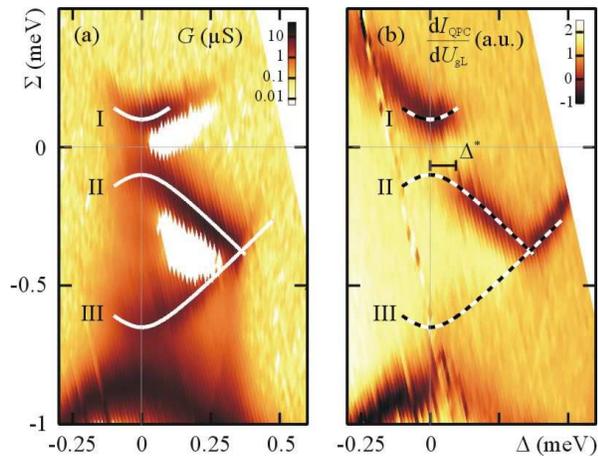, height=6cm}
\end{center}
\vspace*{-0.4cm}
\caption{
(Color online)
(a) Differential conductance ${\rm d}I_{\rm DQD}/{\rm d}U_{\rm SD}$ through the DQD
    corresponding to the current measurement shown in
    Fig.~\ref{fig2}(a) (identical parameters, logarithmic scale).
(b) Corresponding differential transconductance ${\rm d}I_{\rm QPC}/{\rm d}U_{\rm
    gL}$ through a nearby QPC (linear scale).
    For both plots, the model lines are identical to those in
    Fig.~\ref{fig2}. 
}
\label{fig3}
\end{figure}

\begin{figure}[tb]
\begin{center}
\epsfig{file=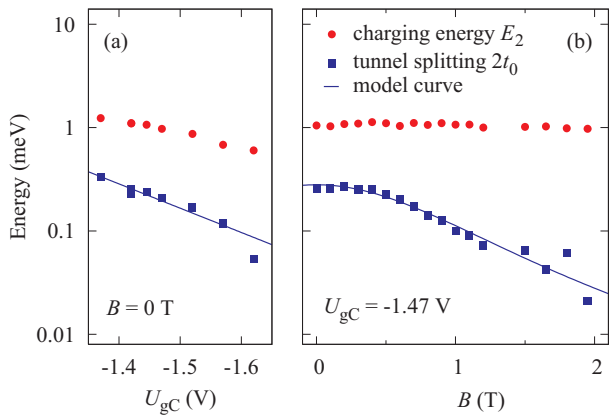, width=8cm}
\end{center}
\vspace*{-0.4cm}
\caption{
(Color online)
(a) Tunnel splitting $2t_0$ of the one--electron DQD
and charging energy for the second electron
$E_{\rm 2}$ in dependence on gate voltage $U_{\rm gC}$.
(b) $E_{\rm 2}$  and $2t_0$ plotted as function of a perpendicular
  magnetic field.
}
\label{fig4}
\end{figure}

Fig.~\ref{fig2}(a)
%
%
displays SET current through the DQD for the region of the stability diagram framed
in Fig.~\ref{fig1}(b) by a rhomboid, but measured at a
source--drain bias of $U_{\rm SD}=-0.75\,\mathrm{mV}$. Here, we use a
new coordinate system reflecting the symmetry properties of the DQD.
The chemical potentials of the QDs marked in Fig.~\ref{fig1}(a) as
$\mu_{\rm L}$ and $\mu_{\rm R}$ are defined as the energies required
to add the first
electron to the respective QD. Our new x--axis measures the interdot asymmetry
$\Delta\equiv(\mu_{\rm R}-\mu_{\rm L})/2$, while the y--axis plots the average
chemical potential of both QDs $\Sigma\equiv(\mu_{\rm R}+\mu_{\rm L})/2$.
Direction and scale of these axes, indicated in Fig.~\ref{fig1}(b) by two
black bars, reflect the capacitances between QDs and
gate electrodes~\cite{hofmann,dodoreview}.

Narrow current features in the stability diagram at zero source--drain bias
expand to regions of finite current for $|U_{\rm SD}|>0$ as
displayed in Fig.~\ref{fig2}(a). Here $\mu_{\rm D}<\mu_{\rm S}$ and for weak
interdot coupling the condition
$\mu_{\rm D}\le\mu_{\rm L}\le\mu_{\rm R}\le\mu_{\rm  S}$
defines a triangular region of finite SET current illustrated in Fig.~\ref{fig2}(b) by dashed
lines~\cite{dodoreview,cmm-spinblock}. The triangle baseline is located on the
$\Sigma$--axis where $\mu_{\rm L}=\mu_{\rm R}$. The other two edges have
slopes ${\rm d}\Sigma/{\rm d}\Delta=\pm 1$ corresponding to either $\mu_{\rm R}=\mu_{\rm S}$ or $\mu_{\rm
L}=\mu_{\rm D}$. They meet at the tip of the triangle with $2 \Delta =\mu_{\rm
S}-\mu_{\rm D} \equiv \left| e U_{\rm SD} \right|$, where $e$ is the electron charge.
The transformation from the coordinates in Fig.~\ref{fig1}(b) to those in
Fig.~\ref{fig2} is based on the geometry of this triangle 
and on the comparison to a reference energy scale provided by $|e U_{\rm SD}|$.

The SET current measurement plotted in Fig.~\ref{fig2}(a) illustrates a
strong interdot coupling situation, where the electronic ground
states of the two QDs hybridize into delocalized states. At
$\mu_{\rm L}=\mu_{\rm R}$ the corresponding chemical potentials $\mu_+$ of the
symmetric ground state and $\mu_-$ of the antisymmetric excited state are
separated by the tunnel splitting $2t_0$. For finite interdot asymmetry
$\Delta$ this energy splitting becomes $\sqrt{(2\Delta)^2+(2t_0)^2}$.

The resulting expected edges of strong (weak) current onset are indicated as
solid (dotted) model lines in Figs.~\ref{fig2}(a) and (b).  Level schemes in
Fig.~\ref{fig2}(b) sketch the alignment of $\mu_+$ and $\mu_-$ compared to the
lead chemical potentials $\mu_{\rm S}$ and $\mu_{\rm D}$ at nearby
intersection points of these lines and the $\Sigma$--axis. For the DQD
containing up to one electron SET is possible in a region
spanned by lines I and III. A first conductance channel opens at line I where
$\mu_+=\mu_{\rm S}$, and a second for $\mu_-=\mu_{\rm S}$ (II). Coulomb blockade
suppresses current for $\mu_+ < \mu_{\rm D}$ (III). 
Therefore, the fourth possible alignment,
$\mu_-=\mu_{\rm D}$, does not appear as a current change. The model
lines I and II result in an anticrossing at zero asymmetry $\Delta=0$ with
tunnel splitting $2t_0 = \left| \mu_- - \mu_+ \right|$.

At even smaller values of $\Sigma$, below line III in Fig.~\ref{fig2}(a), 
the region of high current (dark area) corresponds to the onset of the DQD being
charged with a second electron. Here, Coulomb repulsion and exchange
interaction have to be taken into account, causing a different energy
spectrum.  The in comparison to the background slightly darker graytone in the
region $0 \lesssim \Delta \lesssim |e U_{\rm SD}|$ below line III indicates a
small cotunneling current.

Between lines I and II the antisymmetric excited state of the
DQD is permanently unoccupied because of $\mu_- > \mu_{\rm S}$. SET
is only possible through the delocalized symmetric ground state. In
Fig.~\ref{fig2}(a) we observe current only for small asymmetry
$|\Delta|\lesssim t_0$ between lines  I and II. With increasing asymmetry the
ground state is more and more localized in one of the QDs and  current vanishes
for $\left| \Delta \right| \gg t_0$. A smaller signal for negative
than for positive asymmetry is detected. This broken symmetry hints at
a larger tunnel barrier of the DQD towards its drain contact compared to that on its
source side. In contrast, between lines II and III both electron
states with chemical potentials $\mu_+$ and $\mu_-$
contribute to the current. Here, for $\Delta<0$ energy relaxation
from the left to the right QD competes with the tunnel rate into drain,
additionally hindering SET current through the DQD. For $\Delta\ll0$ strong
localization of the ground state in the right QD near the source contact
causes Coulomb blockade.

In the region spanned by lines II, III, and the $\Sigma$--axis
we observe a strong current signal. Here, the chemical potentials
obey $\mu_{\rm S} \ge \mu_- > \mu_+ \ge \mu_{\rm D}$. For small
asymmetry SET through both then delocalized states contributes to the
current. As the asymmetry $\Delta$ and thereby localization of the two
states in the two separate QDs increases, direct SET through each
of these states is suppressed. For $\Delta \gg t_0$ the electronic states of
the DQD can be approximated by the two single QD ground states with
$\mu_-\simeq\mu_{\rm R}$ and $\mu_+\simeq\mu_{\rm L}$.
In this configuration, resembling the weak coupling limit, an electron can hop
through the double dot starting from the source contact via the right to
the left dot and finally to drain, loosing energy beween the
two dots~\cite{fujisawa-phonons}. This additional energy relaxation process impedes SET and causes a decrease in
total current. 

Fig.~\ref{fig3}(a)
%
%
displays the corresponding differential conductance ${\rm d}I_{\rm DQD}/{\rm d}U_{\rm
SD}$ in logarithmic scale. The tunnel splitting $2t_0$, visible as the
anticrossing of lines I and II, is here even more obvious than for the current
measurement. Note that the model lines in all Figs$.$~\ref{fig2} and
\ref{fig3} are identical. The regions of negative differential conductance
(NDC), plotted in white, reflect the dependence of the tunnel and
relaxation rates on $U_{\rm SD}$. In general, we find that localization
increases with growing $U_{\rm SD}$, in turn causing a slower energy
relaxation between the localized QDs. While this scenario explains NDC
between lines II and III, for the NDC--area above line II in addition higher
order tunnel processes have to be considered involving the excited state
with $\mu_- > \mu_{\rm S}$.

In Fig.~\ref{fig3}(b) the differential transconductance through the QPC ${\rm
d}I_{\rm QPC}/{\rm d}U_{\rm gL}$ is plotted within the same area 
of the stability diagram as the
measurements discussed above. The measured resonances of small signal
(dark areas) are caused by changes of the time--averaged charge inside
the DQD~\cite{field,elzerman}. They follow parts of our model lines
already shown in Figs.~\ref{fig2} and
\ref{fig3}. The dark area at the lower plot edge reflects the onset of
charging the DQD with a second electron. Above this area for each
distinct asymmetry value $\Delta$ only a single charging resonance can
be resolved (the splitted dark feature). Therefore, at this resonance
the charge of the DQD almost discretely switches between zero and one
electron.  At $\Delta \lesssim 0$ charging takes place at the first
possible alignment of the chemical potentials of the DQD and the
source contact ($\mu_+\simeq\mu_{\rm S}$ at model line I), while at an
asymmetry $\Delta^\star>0$ the resonance jumps to line II with
$\mu_-\simeq\mu_{\rm S}$ and, for $2\Delta>e U_{\rm SD}$, follows line
III featuring $\mu_+\simeq\mu_{\rm D}$, where the DQD enters Coulomb
blockade.

To understand the absence of additional charging features along line
III for $2 \Delta < \left| e U_{\rm SD} \right|$ the rates of an
electron entering and escaping the DQD have to be compared.
Information on this is provided by the jump of the charging resonance from
line I to line II. It does not take place at $\Delta =0$
but at a finite positive $\Delta^\ast$. As the asymmetric current
in this region (see above),
this is interpreted such that the tunnel barrier towards the drain contact is
larger than the one on the source side. For $\mu_->\mu_{\rm S}$ and
$\Delta<\Delta^\ast$ an electron tunnels faster into the symmetric
ground state of the DQD than it escapes to the drain contact.
At $\Delta\simeq\Delta^\ast$ both rates are identical, and for
$\Delta>\Delta^\ast$ an electron in the DQD escapes so fast to the drain that
the DQD stays mostly empty along line I. This is caused by the
increasing localisation of the DQD ground state in the left QD near
the drain contact. 

At $\mu_-\simeq\mu_{\rm S}$ (line II) another channel namely via the
excited state opens. The tunnel rate into this state is large, since
for $\Delta > 0$ it is predominantly localized near the source
contact. In comparison, for $\Delta > t_0$ the escape rate is limited
by the product of the relaxation rate between the partly localised QDs
and the tunnel rate from the left QD to the drain reservoir. Thus,
below line II the DQD average charge is approximately one electron.

Fig.~\ref{fig4} 
%
%
plots the tunnel splitting $2t_0$ and the charging energy $E_2$ required
to add a second electron to the DQD as a function of center gate voltage
$U_{\rm gC}$ (a) and perpendicular magnetic field (b). $E_2$
is obtained as the energy difference between the 
onset of current involving one compared to two electrons in a
symmetric DQD. It is taken e$.$g$.$ from Fig.~\ref{fig3}(a) as the distance
between  line I and the center of the dark line at the bottom of this plot.
As we separate the two QDs more and more by increasing the interdot
tunnel barrier (Fig.~\ref{fig4}(a)) both $2t_0$ and $E_2$
decrease. In contrast, with growing magnetic field only $2t_0$ decreases
considerably while $E_2$ stays almost unaltered. Note that $E_2$ primarily
depends on the distance of the centers of charge of both QDs. In comparison, the tunnel
rate is additionally influenced by the overall potential geometry governing the
overlap of the QDs localized electronic states. Clearly, $U_{\rm gC}$ alters the
electrostatic potential geometry whereas $B$ mainly leads to a compression of
the electronic states with almost constant mean distance. Thus $E_2(B)$
remains almost unchanged.

Using the WKB--method the tunnel splitting of the DQD is found to be
approximately $2t_0 \simeq 2E_0/\pi$ $\exp(-\sqrt{2m^\ast\Phi}\,d/2\hbar)$,
where $E_0$ is the mean ground state energy of both QDs, $m^\ast$ the
effective electron mass, and where $d$ and $\Phi$ are the effective width and
amplitude of a quartic tunnel barrier potential. An exponential decrease of
$2t_0$ at increasing $|U_{\rm gC}|$, as indicated by a solid line in
Fig.~\ref{fig4}(a), suggests a nearly linear increase of $\sqrt{\Phi} \, d$.

Fig.~\ref{fig4}(b) displays $2t_0$ and $E_2$ at a fixed voltage $U_{\rm
gC}=-1.47\,\mathrm{V}$ in dependence of magnetic field $B$. The tunnel
splitting remains constant at $2t_0 \simeq 260  \, \mu\mathrm{eV}$ for $B
\lesssim 0.4 \, \mathrm{T}$, but decreases for stronger $B$.  At constant mean
distance, we assume each QD to extend over the Fock--Darwin length scale
$l_{\rm QD}(B) \equiv \sqrt{\hbar / \omega_c m^\ast } /
\sqrt[4]{1+ 4\Omega^2/\omega_c^2} $~\cite{fock,dots}. Here $\Omega$
characterizes the parabolic confinement and $\omega_c = e B /
m^\ast$. The WKB--formula then results in a model curve (solid line) showing
qualitative good agreement with our data. 
For a quantitative analysis the actual
overlap of the wavefunctions within a realistic potential
had to be considered.

In conclusion, we have directly observed the coherent quantum mechanical interdot tunnel
coupling of a one--electron DQD employing quasi--static transport measurements.
At finite source--drain bias the delocalized electronic eigenstates of the
strongly coupled DQD generate a distinct pattern in the stability diagram visible
in current, conductance, and average charge on the DQD.
In all three quantities, the tunnel splitting is immediately visible as a
clear anticrossing and can be quantified after a coordinate transformation.
To tune the tunnel splitting we modify gate voltages or a magnetic field
perpendicular to the 2DES. We propose a simple model and find our data in
qualitative agreement.

We thank R$.$ Blick, H$.$ Lorenz, U$.$ Hartmann, and F$.$~Wilhelm for
valuable discussions, and S$.$ Manus for expert technical help, as well as
the Deut\-sche For\-schungs\-ge\-mein\-schaft and the
BMBF for support. 
A$.$ K$.$ H$.$ thanks the German Nat.
Acad. Foundation for support. 


\bibliographystyle{apsrev}

\end{document}